\documentclass[aps,prl,twocolumn,groupedaddress]{revtex4-1}

\usepackage{graphicx}

\begin{document}


\title{Identification of Patient Zero in Static and Temporal Networks - Robustness and Limitations}



\author{Nino Antulov-Fantulin}
\affiliation{Computational Biology and Bioinformatics Group, Division of Electronics, Rudjer Bo\v{s}kovi\'{c} Institute, Zagreb 10000, Croatia }%

\author{Alen Lan\v{c}i\'{c}}
\affiliation{Faculty of Science, Department of Mathematics,\\ University of Zagreb, Zagreb 10000, Croatia}%

\author{Tomislav \v{S}muc}
\affiliation{Computational Biology and Bioinformatics Group, Division of Electronics, Rudjer Bo\v{s}kovi\'{c} Institute, Zagreb 10000, Croatia}%

\author{Hrvoje \v{S}tefan\v{c}i\'{c}}
\affiliation{Theoretical Physics Division, Rudjer Bo\v{s}kovi\'{c} Institute, Zagreb 10000, Croatia}%
\affiliation{Catholic University of Croatia, Zagreb, Croatia}%

\author{Mile \v{S}iki\'{c}}
\affiliation{Faculty of Electrical Engineering and Computing, Department of Electronic Systems and Information Processing, University of Zagreb, Zagreb 10000, Croatia}%
\email[Corresponding mail:]{ mile.sikic@fer.hr}
\affiliation{Bioinformatics Institute, A*STAR, Singapore 138671, Republic of Singapore}%


\date{\today}

\begin{abstract}
Detection of patient-zero can give new insights to the epidemiologists about the nature of first transmissions into a population.
In this paper, we study the statistical inference problem of detecting the source of epidemics from a snapshot of spreading on an arbitrary network structure. 
By using exact analytic calculations and Monte Carlo estimators, we demonstrate the detectability limits for the SIR model, which primarily depend on the spreading process characteristics. 
Finally, we demonstrate the applicability of the approach in a case of a simulated sexually transmitted infection spreading over an empirical temporal network of sexual interactions.
\end{abstract}


\maketitle

\section*{Introduction}
One of the most prevalent types of dynamic processes of public interest characteristic for the real-life complex networks are contagion processes \cite{VespignaniNPhys2011, GLEaMviz, MobilityProxies, ColizzaReactionDiffusionProcess, VespignaniNPhys2011, Castellano10, EpidScaleFree, FastSIR}.
Epidemiologists detect the epidemic source or the patient-zero either by analysing the temporal genetic evolution of virus strains \cite{PNASPandemic1918, AntiGeneicEvolution, SourcePhylogenetic}, which can be time-demanding or try to do a contact backtracking \cite{AIDS_patientZero} from the available observed data. 
However, in cases where the information on the times of contact is unknown or incomplete or the infection is asymptomatic or subclinical the backtracking method is no longer adequate. 
Due to its practical aspects and theoretical importance, the epidemic source detection problem on contact networks has recently gained a lot of attention in the complex network science community. This has led to the development of many different source detection estimators for static networks, which vary in their assumptions on the network structure (locally tree-like) or on the spreading process compartmental models (SI, SIR) \cite{Zaman1, RegularTreeSI_11, MultipleObersvations, Jaccard_7, pinto2012, DMP_0, BBP, SourceDetectAvgTopK, SourceEigenCen_5, Brockmann} or both.

In the case of the SIR model (Susceptible-Infected-Recovered) there are two different approaches. Zhu et. al. proposed a sample path counting approach \cite{Jaccard_7}, where they proved that the source node minimizes the maximum distance (Jordan centrality) to the infected nodes on infinite trees. Lokhov et. al. used a dynamic message-passing algorithm (DMP) for the SIR model to estimate the probability that a given node produces the observed snapshot. They use a mean-field-like approximation (node independence approximation) and an assumption of a tree-like contact network to compute the source likelihoods \cite{DMP_0}. Altarelli. et. al. remove the independence assumption and use the message passing method with an assumption of a tree-like contact network to estimate the source \cite{BBP}. In our study, we drop all the network structure and node independence assumptions and analyse the source probability estimators for general compartmental models. 
The main contributions of our paper are the following:  

(i) we developed the analytic combinatoric, as well as the Monte-Carlo methods (Direct and Soft Margin) for determining exact and approximate source probability distribution, and have also produced the benchmark solutions on the 4-connected regular lattice structure;

(ii) we measured the source detectability by using the normalized Shannon entropy of the estimated source probability distribution for each of the source detection problems, and have observed the existence of some highly detectable, as well as some highly undetectable regimes for the SIR and other spreading models. We notice that the detectability primarily depends on the spreading process characteristics;

(iii) using the simulations of the sexually transmitted infection (STI) on a realistic time interval of 200 days on an empirical temporal network of sexual contacts we demonstrate the robustness of the Soft Margin source estimator.

\section*{Methods}
In a general case, the contact-network during an epidemic process can be temporal and weighted, but we first concentrate our analysis on a static undirected and non-weighted network $G=(V,E)$, where $V$ denotes a set of nodes and $E$ denotes a set of edges. The random binary vector $\vec{R}$ indicates which nodes got infected up to a certain time $T$. For the contagion model, we use the SIR model with the simultaneous updates in time described by the probability $p$ that an infected node infects a susceptible neighbour node in one discrete step and the probability $q$ that an infected node recovers in one discrete step.
We observe one epidemic realization $\vec{r}_*$ of $\vec{R}$ at a time $T$ of the SIR process $(p,q,T)$ on a network $G$ and want to calculate the source posterior probabilities $P(\Theta = \theta_i|\vec{R}=\vec{r}_*)$. We have developed two complementary approaches that can provide exact posterior probability distributions over nodes in the spreading realization $\vec{r}_{*}$ via the Bayesian approach: the direct Monte-Carlo approach and analytical combinatoric approach.

Using \textbf{the direct Monte-Carlo approach}, for each potential source node $i$ (infected node in the realization $\vec{r}_{*}$), a large number $n$ of epidemic spreading simulations with maximum duration $T$ is performed with $i$ as an epidemic source. The number of simulations $n_i$ which coincides with the realization $\vec{r}_{*}$ is recorded. To cut down on the extensive calculation required for the Monte-Carlo simulations, we employ a pruning mechanism (no errors introduced), stopping the simulations at $t<T$ if the current simulation realization has infected a node which is not infected in $\vec{r}_{*}$. The probability of the node $i$ being the source of the epidemic is then calculated as $P(\Theta = \theta_i|\vec{R}=\vec{r}_*) = n_i / \sum_j n_j$.  The statistical significance of the direct Monte-Carlo results are controlled with the convergence conditions. For more information, see SI section 2.

\begin{figure}
\includegraphics[scale=0.45]{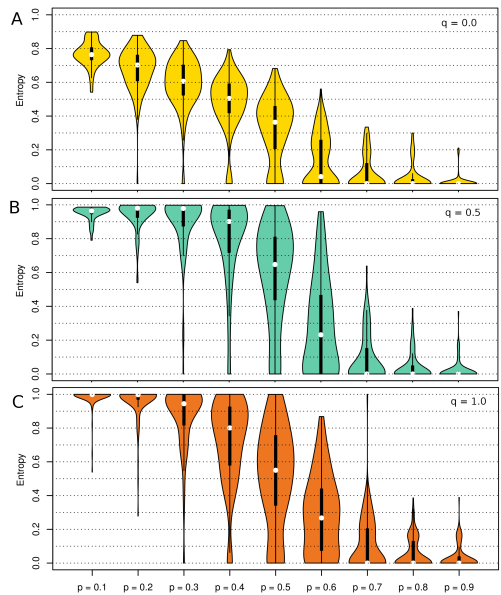}
\caption{
Plots A, B and C: Box plots depicting distribution of entropy values (H) of source probability distributions for a number of randomly generated spreading realizations across with different $(p,q)$ parameters on the 4-connected lattice: $N = 30 \times 30$ nodes with $T = 5$, calculated by the direct Monte-Carlo method with $10^6-10^8$ simulations per source.}
\label{fig:IllPoseCase}
\end{figure}

An alternative approach, \textbf{the analytical combinatoric approach} assigns to each node of degree $n$ a generating function which is maximally $(n+1)$-dimensional, which captures the events of node first infection and infection spreading through its edges at specific times. Then, by multiplication of the generating functions of all the infected nodes from a realization, we are able to merge all contributions together and get the source probability distribution. In the SI section 1, along with the detailed description of analytical combinatoric method, we demonstrate the correspondence between the direct Monte-Carlo and analytical combinatorics. The detailed description of analytical combinatoric method can be found in SI section 1. A serious disadvantage of the analytical method is that the calculations become prohibitively intricate in the case of non tree-like configurations.

We have generated a series of benchmark cases on a 4-connected lattice ($N=30 \times 30$), for which we have calculated the probability distributions over the potential source candidates using the direct Monte-Carlo estimator (see the SI section 4). 
The source \textbf{detectability} $D(\vec{r}_{*})=1-H(\vec{r}_{*})$, is characterized via the normalized Shannon entropy $H$ (normalization by entropy of uniform distribution) of the calculated probability distribution $P(\Theta = \theta_i|\vec{R}=\vec{r}_*)$.

\begin{figure}
\includegraphics[scale=0.45]{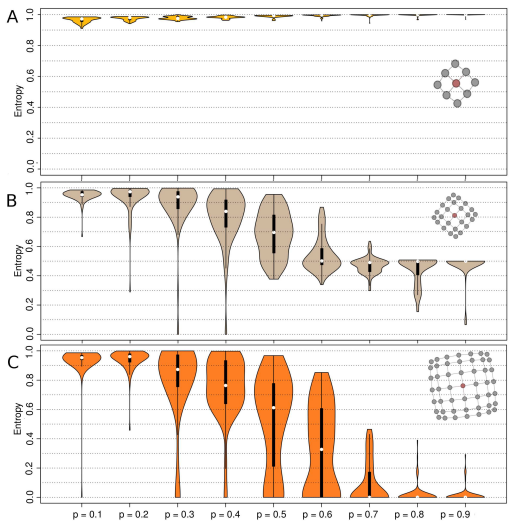}
\caption{
Plots A, B and C: Box plots depicting distribution of entropy values (H) of source probability distributions for a number of realizations starting from the central node denoted with red colour on the 4-connected lattice with different sizes ($3\times 3$, $5\times5$ and $7\times7$) with the SIR model for $q=0.5$, $T = 5$ and different $p$ values, calculated with the Soft Margin method with the ($10^4$ - $10^6$) simulations per source and adaptive $a$ chosen from the convergence condition.
}
\label{fig:SoftGrids}
\end{figure}

Results depicting distributions of $H$ for different parts of the SIR parameter space for the regular lattice are given in Figure \ref{fig:IllPoseCase} plots A, B and C. 
Figures show qualitatively the same detectability behaviour across $p$ parameter, for different values of parameter $q$. It is important to observe the existence of three different regions: low detectability-high entropy region $(p<0.2)$, intermediate detectability-intermediate entropy region $(0.2<p<0.7)$ and high detectability-low entropy region $(p>0.7)$.
We observe that the detectability transition is still present even for different spreading models (SI, ISS, IC) and we observe the interplay of the network size and stopping time $T$ on the detectability (see SI section 10 and Figure \ref{fig:SoftGrids}). In Figure \ref{fig:SoftGrids}, plot A, we observe that in a regime, when the network size restricts the epidemic spreading but not the epidemic itself via it’s natural evolution characterized by the parameters $(p, q)$ or stopping time $T$, the entropy is high as the realizations from different sources are almost identical.

As application of direct Monte-Carlo and analytical combinatoric approaches becomes prohibitively expensive for realistic network sizes, we formulate an estimator which is much more efficient in approximating the true underlying source probability distribution for the particular epidemic spread.
We continue with the definition of the \textbf{Soft-Margin} estimator, a generalization of the Monte-Carlo inference method, in which direct Monte-Carlo method represents a limiting case.
In order to proceed we first need to introduce some useful definitions. 
The random binary vector $\vec{R}_{\theta}$ describes the outcome of epidemic process and sample vectors: $\left\lbrace \vec{r}_{\theta,1}, ..., \vec{r}_{\theta,n} \right\rbrace$ describe $n$ independent outcomes of that process. Each sample vector $\vec{r}_{\theta,i}$ is obtained using the Monte Carlo simulation of the contagion process with the $\theta$ as the source. We measure the similarity between vectors $\vec{r_1}$ and $\vec{r_2}$ by the Jaccard similarity function $\varphi:(R^N \times R^N) \to [0,1]$ calculated as the ratio of the size of the interaction of set of infected nodes in $\vec{r_1}$, $\vec{r_2}$ and the size of their union. 
The random variable $\varphi(\vec{r_*}, \vec{R_{\theta}} )$ measures the similarity between a fixed realization $\vec{r_*}$ and a random vector realization that comes from $SIR$ process with the source $\theta$. 
The empirical cumulative distribution function of the $n$ samples from the random variable $\varphi(\vec{r_*}, \vec{R_{\theta}})$ is denoted $\hat{F}_{\theta}(x)$, where $x$ is the value of the similarity variable. By taking the derivative of $\hat{F}_{\theta}(x)$, we get the PDF estimate:
\begin{equation}
\label{eq:PDF}
\hat{f}_{\theta}(x) = \frac{d}{dx} \hat{F_{\theta}}(x) = \frac{1}{n} \sum_{i=1}^n \delta \left( x- \varphi(\vec{r_*}, \vec{r}_{\theta,i}) \right),
\end{equation}
where $\delta(x)$ denotes the Dirac delta distribution. 
Having defined the PDF for the observed similarities $\hat{f}_{\theta}(x)$, we can now define the main Soft-Margin inference expression as:
\begin{equation}
\hat{P}(\vec{R} = \vec{r_*} | \Theta = \theta) = 
\int_0^1 w_a(x) \hat{f}_{\theta}(x) \mathrm{d}x,
\end{equation}
where $w_a(x)$ is a weighting function. We use the following Gaussian weighting form: $w_a(x) = exp(-(x-1)^2/a^2)$.
In the limit where the parameter $a \to 0$, we obtain the direct Monte-Carlo likelihood estimation. 
For cases when the parameter $a > 0$, we obtain an estimator which estimates the likelihood by using the weighting function $w_a(x)$ to accept contributions from realizations whose similarity to observed realization is less than $1$. 
Using the property of delta distribution, we simplify the expression for the Soft Margin estimator to (for more details see SI section 5):
\begin{equation}
\hat{P}(\vec{R} = \vec{r}_* | \Theta = \theta)= \frac{1}{n} \sum_{i=1}^n exp \left( \frac{-(\varphi(\vec{r_*}, \vec{r}_{\theta,i})-1)^2}{a^2} \right).
\end{equation}
Note, that alternative view on the Soft margin estimator is the non-parametric density estimation with the Gaussian kernels \cite{KernelEstimation}. 
Finally, we do not need to set the Soft Margin width parameter $a$ in advance. After we calculate the estimated PDF for every potential source $\hat{F_{\theta}}(x)$, we can choose the parameter $a$ as the infimum of the set of parameters for which the PDFs have converged. 
The implementation details, time complexity analysis and pruning mechanism for the Soft Margin estimator can be found in the SI sections 5, 6 and 7.

\section*{Results}
\begin{figure}
\includegraphics[scale=0.5]{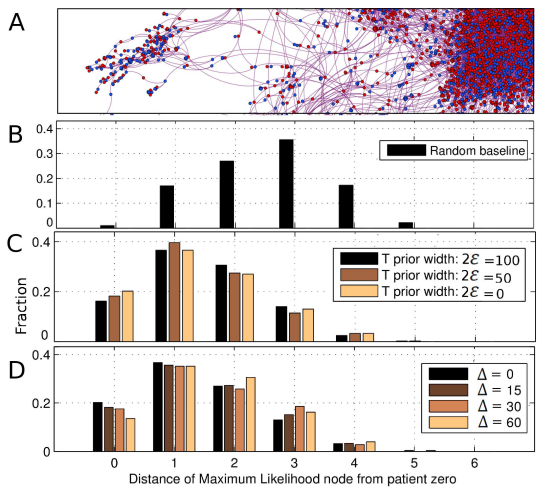}
\caption{ 
Plot A: Visualization of a part of the aggregated empirical temporal network of sexual contacts in Brazil \cite{RochaHolme}. 
In plots B,C and D the performance is measured as the fraction of 500 experiments with specific graph distance of the maximum likelihood candidate to the true source.
The average execution time of a single experiment to calculate source probability distribution over all potential candidates was around 12 seconds (on 50 cpu cores) with $20000$ STI simulations per node.
Plot B: The baseline performance of a random estimator, which uniformly assigns likelihood to potential nodes.
Plot C: The influence of prior knowledge about initial outbreak moment $[t_0 - \epsilon, t_0 + \epsilon]$ of the outbreak on performance.
Plot D: The influence of randomized temporal ordering of interactions within $\Delta$ days, with $\epsilon = 0$ (we know the starting time $t_0$) on performance. 
}
\label{fig:temporal}
\end{figure}

We now demonstrate the applicability of our inference framework to detect the source of the simulated STI epidemic spreading in an empirical temporal network of sexual contacts in Brazil (see Figure \ref{fig:temporal} plot A). This publicly available dataset \cite{RochaHolme} was obtained from Brazilian Internet community and is used as an approximation of temporal sexual contacts. The dataset (see SI section 8) consists out of the triplets $(v_i,v_j,t)$, which represents the event that the nodes $v_i$ and $v_j$ had a sexual interaction at a time $t$. First $1000$ days in original dataset are discarded due to the transient period with sparse encounters \cite{RochaHolme} and therefore all temporal moments are measured relative to day $1000$, as have done the authors in the original study \cite{RochaHolme}. For our temporal network, we use the SIR model $(p = 0.3, q = 0.01)$ for STI. The upper limit of the transmission probability for the STI that was previously used on this contact network is $p=0.3$ \cite{RochaHolme}. The recovery parameter $q=0.01$ represents a disease with the mean recovery of 100 days.

Note that here the calculation of exact source probability distributions is computationally too demanding for both the direct Monte-Carlo and the analytical combinatoric method. Therefore, we use the Soft Margin estimator with the smallest width $a$ for which the ML node probability estimate converged. Our experiments consist of two parts: (i) simulation of STI spreading through a temporal network of sexual contacts and (ii) detection of the patient zero from the observed process. 

In order to demonstrate applicability of the approach in realistic conditions, we introduce uncertainty in the epidemic starting time $t_0$, and later on also with respect to node states in observed epidemic realization. Note, that uncertainties in $(p,q)$ parameters can also be relaxed by marginalization procedure (see SI section 5).
The relaxation of knowing the starting point of the epidemic $t_0$ is done by using the marginalization over time, sampling over all possible starting points $t_0$ from a uniform probability distribution over $[t_0 - \epsilon, t_0 + \epsilon]$, $2\epsilon = \{ 0, 50, 100 \}$ days. 
In Figure \ref{fig:temporal} plot C, we show the summary results from 500 independent experiments, when the starting time $t_0$ was chosen from the interval of $[100-200]$ days, the end of the epidemic was set to the day $t=300$ and using different uniform priors ($\epsilon$) for the moment $t_0$.  
Using the uniform uncertainty of $\epsilon=50$ days, we can still detect the source within its first neighbourhood (distance 0 and 1 from the source) in approximately 60\% of experiments. These results are of great practical importance, since in reality we do not know the exact starting times, but rather only an upper and a lower bound on starting point.  

Next, we demonstrate how the uncertainty in the temporal orderings of interactions within a time window of the length $\Delta$ affects the performance of source detection. We use a randomization algorithm which permutes time stamps inside of a bin of $\Delta$ days from the start to the end of the contact interaction network in a non-overlapping way. From Figure \ref{fig:temporal}, plot D, we observe that higher uncertainty in orderings (higher $\Delta$) reduces the detectability of the source of infection. However, the estimation framework is robust to small-scale interaction noise.

We have also shown that our Soft Margin algorithm estimates source probabilities with much higher precision than other estimators (Jordan and DMP estimator) on benchmark cases by comparing the results against the direct Monte-Carlo source probability estimations on regular lattice (see SI section 3).
Results for source detection for different values of $(p,q)$ parameters and for the case when only a random subset of the node states is observed can be found in SI section 9. 

\section*{Discussion}
The assumption about missing dynamic information about times of infection or recovery in our case study seems rather plausible for two realistic cases: STI infections and computer viruses. Many STIs generate silent epidemics since many of them are unrecognized, asymptomatic or subclinical as the pathogens are being transmitted from patients with mild or totally absent symptoms. A large number of people with STIs: chlamydia \cite{GonnChly}, gonorrhea \cite{GonnChly}, Human papillomavirus and other show mild or no symptoms at all. The second motivation comes from silent spreading of a certain class of computer viruses and worms through computer networks which become active simultaneously on a specific date.
Unlike other approaches \cite{Zaman1, RegularTreeSI_11, MultipleObersvations, Jaccard_7, pinto2012, DMP_0, BBP, SourceDetectAvgTopK, SourceEigenCen_5, Brockmann}, we identified different source detectability regimes and our methodology is applicable to arbitrary network structures, and is limited solely by the ability to computationally produce realizations of the particular contagion process.

\begin{acknowledgments}
\section*{Acknowledgments}
The authors would like to thank: Professor Dirk Brockmann for valuable discussions about the epidemic source detection problems during internship of NAF at the Robert Koch-Institute in Berlin, Germany. For proofreading the manuscript we would like to thank: Vinko Zlati\'{c}, Sebastian Krause and Ana Bulovi\'{c}. The work is financed in part by: the Croatian Science Foundation under 
the project number I-1701-2014, the EU-FET project MULTIPLEX under the grant number 317532 and the FP7-REGPOT-2012-2013-1 InnoMol project under the grant number 316289.\\
\textbf{Disclaimer}: The authors of this study used the published existing dataset of sexual contacts in high-end prostitution because it contains valuable and rarely available information on temporal network of contacts serving as pathways of STD spreading. It is important to note that the use of this dataset does not reflect the authors' views, opinions and attitudes on prostitution and it does not in any way imply that the authors support the activities documented in the dataset or the way the data were gathered.
\end{acknowledgments}

\bibliography{myBib2.bib}

\begin{thebibliography}{25}
\expandafter\ifx\csname natexlab\endcsname\relax\def\natexlab#1{#1}\fi
\expandafter\ifx\csname bibnamefont\endcsname\relax
  \def\bibnamefont#1{#1}\fi
\expandafter\ifx\csname bibfnamefont\endcsname\relax
  \def\bibfnamefont#1{#1}\fi
\expandafter\ifx\csname citenamefont\endcsname\relax
  \def\citenamefont#1{#1}\fi
\expandafter\ifx\csname url\endcsname\relax
  \def\url#1{\texttt{#1}}\fi
\expandafter\ifx\csname urlprefix\endcsname\relax\def\urlprefix{URL }\fi
\providecommand{\bibinfo}[2]{#2}
\providecommand{\eprint}[2][]{\url{#2}}

\bibitem[{\citenamefont{Vespignani}(2012)}]{VespignaniNPhys2011}
\bibinfo{author}{\bibfnamefont{A.}~\bibnamefont{Vespignani}},
  \bibinfo{journal}{Nat Phys} \textbf{\bibinfo{volume}{8}}, \bibinfo{pages}{32}
  (\bibinfo{year}{2012}), ISSN \bibinfo{issn}{1745-2473},
  \urlprefix\url{http://dx.doi.org/10.1038/nphys2160}.

\bibitem[{\citenamefont{Broeck et~al.}(2011)\citenamefont{Broeck, Gioannini,
  Goncalves, Quaggiotto, Colizza, and Vespignani}}]{GLEaMviz}
\bibinfo{author}{\bibfnamefont{W.}~\bibnamefont{Broeck}},
  \bibinfo{author}{\bibfnamefont{C.}~\bibnamefont{Gioannini}},
  \bibinfo{author}{\bibfnamefont{B.}~\bibnamefont{Goncalves}},
  \bibinfo{author}{\bibfnamefont{M.}~\bibnamefont{Quaggiotto}},
  \bibinfo{author}{\bibfnamefont{V.}~\bibnamefont{Colizza}}, \bibnamefont{and}
  \bibinfo{author}{\bibfnamefont{A.}~\bibnamefont{Vespignani}},
  \bibinfo{journal}{BMC Infectious Diseases 11, 37 (2011)}
  (\bibinfo{year}{2011}),
  \urlprefix\url{http://www.biomedcentral.com/1471-2334/11/37}.

\bibitem[{\citenamefont{Tizzoni et~al.}(2014)\citenamefont{Tizzoni, Bajardi,
  Decuyper, Kon Kam~King, Schneider, Blondel, Smoreda, González, and
  Colizza}}]{MobilityProxies}
\bibinfo{author}{\bibfnamefont{M.}~\bibnamefont{Tizzoni}},
  \bibinfo{author}{\bibfnamefont{P.}~\bibnamefont{Bajardi}},
  \bibinfo{author}{\bibfnamefont{A.}~\bibnamefont{Decuyper}},
  \bibinfo{author}{\bibfnamefont{G.}~\bibnamefont{Kon Kam~King}},
  \bibinfo{author}{\bibfnamefont{C.~M.} \bibnamefont{Schneider}},
  \bibinfo{author}{\bibfnamefont{V.}~\bibnamefont{Blondel}},
  \bibinfo{author}{\bibfnamefont{Z.}~\bibnamefont{Smoreda}},
  \bibinfo{author}{\bibfnamefont{M.~C.} \bibnamefont{González}},
  \bibnamefont{and} \bibinfo{author}{\bibfnamefont{V.}~\bibnamefont{Colizza}},
  \bibinfo{journal}{PLoS Comput Biol} \textbf{\bibinfo{volume}{10}},
  \bibinfo{pages}{e1003716} (\bibinfo{year}{2014}),
  \urlprefix\url{http://dx.doi.org/10.1371%2Fjournal.pcbi.1003716}.

\bibitem[{\citenamefont{Colizza et~al.}(2007)\citenamefont{Colizza, Satorras,
  and Vespignani}}]{ColizzaReactionDiffusionProcess}
\bibinfo{author}{\bibfnamefont{V.}~\bibnamefont{Colizza}},
  \bibinfo{author}{\bibfnamefont{R.~P.} \bibnamefont{Satorras}},
  \bibnamefont{and}
  \bibinfo{author}{\bibfnamefont{A.}~\bibnamefont{Vespignani}},
  \bibinfo{journal}{Nat Phys} \textbf{\bibinfo{volume}{3}},
  \bibinfo{pages}{276} (\bibinfo{year}{2007}),
  \urlprefix\url{http://dx.doi.org/10.1038/nphys560}.

\bibitem[{\citenamefont{Castellano and Pastor-Satorras}(2010)}]{Castellano10}
\bibinfo{author}{\bibfnamefont{C.}~\bibnamefont{Castellano}} \bibnamefont{and}
  \bibinfo{author}{\bibfnamefont{R.}~\bibnamefont{Pastor-Satorras}},
  \bibinfo{journal}{Phys Rev Lett} \textbf{\bibinfo{volume}{105}},
  \bibinfo{pages}{218701} (\bibinfo{year}{2010}).

\bibitem[{\citenamefont{Pastor-Satorras and Vespignani}(2001)}]{EpidScaleFree}
\bibinfo{author}{\bibfnamefont{R.}~\bibnamefont{Pastor-Satorras}}
  \bibnamefont{and}
  \bibinfo{author}{\bibfnamefont{A.}~\bibnamefont{Vespignani}},
  \bibinfo{journal}{Phys. Rev. Lett.} \textbf{\bibinfo{volume}{86}},
  \bibinfo{pages}{3200} (\bibinfo{year}{2001}).

\bibitem[{\citenamefont{Antulov-Fantulin
  et~al.}(2013{\natexlab{a}})\citenamefont{Antulov-Fantulin, Lancic, Stefancic,
  and Sikic}}]{FastSIR}
\bibinfo{author}{\bibfnamefont{N.}~\bibnamefont{Antulov-Fantulin}},
  \bibinfo{author}{\bibfnamefont{A.}~\bibnamefont{Lancic}},
  \bibinfo{author}{\bibfnamefont{H.}~\bibnamefont{Stefancic}},
  \bibnamefont{and} \bibinfo{author}{\bibfnamefont{M.}~\bibnamefont{Sikic}},
  \bibinfo{journal}{Information Sciences} \textbf{\bibinfo{volume}{239}},
  \bibinfo{pages}{226 } (\bibinfo{year}{2013}{\natexlab{a}}), ISSN
  \bibinfo{issn}{0020-0255},
  \urlprefix\url{http://dx.doi.org/10.1016/j.ins.2013.03.036}.

\bibitem[{\citenamefont{Worobey et~al.}(2014)\citenamefont{Worobey, Han, and
  Rambaut}}]{PNASPandemic1918}
\bibinfo{author}{\bibfnamefont{M.}~\bibnamefont{Worobey}},
  \bibinfo{author}{\bibfnamefont{G.-Z.} \bibnamefont{Han}}, \bibnamefont{and}
  \bibinfo{author}{\bibfnamefont{A.}~\bibnamefont{Rambaut}},
  \bibinfo{journal}{Proceedings of the National Academy of Sciences} pp.
  \bibinfo{pages}{201324197+} (\bibinfo{year}{2014}), ISSN
  \bibinfo{issn}{1091-6490},
  \urlprefix\url{http://dx.doi.org/10.1073/pnas.1324197111}.

\bibitem[{\citenamefont{Du et~al.}(2012)\citenamefont{Du, Dong, Lan, Peng, Wu,
  Zhang, Huang, Wang, Wang, Guo et~al.}}]{AntiGeneicEvolution}
\bibinfo{author}{\bibfnamefont{X.}~\bibnamefont{Du}},
  \bibinfo{author}{\bibfnamefont{L.}~\bibnamefont{Dong}},
  \bibinfo{author}{\bibfnamefont{Y.}~\bibnamefont{Lan}},
  \bibinfo{author}{\bibfnamefont{Y.}~\bibnamefont{Peng}},
  \bibinfo{author}{\bibfnamefont{A.}~\bibnamefont{Wu}},
  \bibinfo{author}{\bibfnamefont{Y.}~\bibnamefont{Zhang}},
  \bibinfo{author}{\bibfnamefont{W.}~\bibnamefont{Huang}},
  \bibinfo{author}{\bibfnamefont{D.}~\bibnamefont{Wang}},
  \bibinfo{author}{\bibfnamefont{M.}~\bibnamefont{Wang}},
  \bibinfo{author}{\bibfnamefont{Y.}~\bibnamefont{Guo}}, \bibnamefont{et~al.},
  \bibinfo{journal}{Nature Communications} \textbf{\bibinfo{volume}{3}},
  \bibinfo{pages}{709} (\bibinfo{year}{2012}),
  \urlprefix\url{http://dx.doi.org/10.1038/ncomms1710}.

\bibitem[{\citenamefont{Volz and Frost}(2013)}]{SourcePhylogenetic}
\bibinfo{author}{\bibfnamefont{E.~M.} \bibnamefont{Volz}} \bibnamefont{and}
  \bibinfo{author}{\bibfnamefont{S.~D.~W.} \bibnamefont{Frost}},
  \bibinfo{journal}{PLoS Comput Biol} \textbf{\bibinfo{volume}{9}},
  \bibinfo{pages}{e1003397} (\bibinfo{year}{2013}),
  \urlprefix\url{http://dx.doi.org/10.1371%2Fjournal.pcbi.1003397}.

\bibitem[{\citenamefont{Auerbach et~al.}(1984)\citenamefont{Auerbach, Darrow,
  Jaffe, and Curran}}]{AIDS_patientZero}
\bibinfo{author}{\bibfnamefont{D.~M.} \bibnamefont{Auerbach}},
  \bibinfo{author}{\bibfnamefont{W.~W.} \bibnamefont{Darrow}},
  \bibinfo{author}{\bibfnamefont{H.~W.} \bibnamefont{Jaffe}}, \bibnamefont{and}
  \bibinfo{author}{\bibfnamefont{J.~W.} \bibnamefont{Curran}},
  \bibinfo{journal}{The American Journal of Medicine}
  \textbf{\bibinfo{volume}{76}}, \bibinfo{pages}{487 } (\bibinfo{year}{1984}),
  ISSN \bibinfo{issn}{0002-9343},
  \urlprefix\url{http://dx.doi.org/10.1016/0002-9343(84)90668-5}.

\bibitem[{\citenamefont{Shah and Zaman}(2010)}]{Zaman1}
\bibinfo{author}{\bibfnamefont{D.}~\bibnamefont{Shah}} \bibnamefont{and}
  \bibinfo{author}{\bibfnamefont{T.}~\bibnamefont{Zaman}}, in
  \emph{\bibinfo{booktitle}{Proceedings of the ACM SIGMETRICS international
  conference on Measurement and modeling of computer systems}}
  (\bibinfo{publisher}{ACM}, \bibinfo{address}{New York, NY, USA},
  \bibinfo{year}{2010}), SIGMETRICS '10, pp. \bibinfo{pages}{203--214}, ISBN
  \bibinfo{isbn}{978-1-4503-0038-4},
  \urlprefix\url{http://doi.acm.org/10.1145/1811039.1811063}.

\bibitem[{\citenamefont{Dong et~al.}(2013)\citenamefont{Dong, Zhang, and
  Tan}}]{RegularTreeSI_11}
\bibinfo{author}{\bibfnamefont{W.}~\bibnamefont{Dong}},
  \bibinfo{author}{\bibfnamefont{W.}~\bibnamefont{Zhang}}, \bibnamefont{and}
  \bibinfo{author}{\bibfnamefont{C.~W.} \bibnamefont{Tan}}, in
  \emph{\bibinfo{booktitle}{Proceedings of the IEEE Intl. Symp. on Information
  Theory 2013}} (\bibinfo{year}{2013}),
 \urlprefix\url{http://dx.doi.org/10.1109/ISIT.2013.6620711}.


\bibitem[{\citenamefont{Wang et~al.}(2014)\citenamefont{Wang, Dong, Zhang, and
  Tan}}]{MultipleObersvations}
\bibinfo{author}{\bibfnamefont{Z.}~\bibnamefont{Wang}},
  \bibinfo{author}{\bibfnamefont{W.}~\bibnamefont{Dong}},
  \bibinfo{author}{\bibfnamefont{W.}~\bibnamefont{Zhang}}, \bibnamefont{and}
  \bibinfo{author}{\bibfnamefont{C.~W.} \bibnamefont{Tan}},
  \bibinfo{journal}{SIGMETRICS Perform. Eval. Rev.}
  \textbf{\bibinfo{volume}{42}}, \bibinfo{pages}{1} (\bibinfo{year}{2014}),
  ISSN \bibinfo{issn}{0163-5999},
  \urlprefix\url{http://doi.acm.org/10.1145/2637364.2591993}.

\bibitem[{\citenamefont{Zhu and Ying}(2014)}]{Jaccard_7}
\bibinfo{author}{\bibfnamefont{K.}~\bibnamefont{Zhu}} \bibnamefont{and}
  \bibinfo{author}{\bibfnamefont{L.}~\bibnamefont{Ying}}, in
  \emph{\bibinfo{booktitle}{IEEE/ACM Transactions on Networking}} (\bibinfo{year}{2014}), 
vol.PP, no.99, pp.1, 
\urlprefix\url{http://dx.doi.org/10.1109/TNET.2014.2364972}.

\bibitem[{\citenamefont{Pinto et~al.}(2012)\citenamefont{Pinto, Thiran, and
  Vetterli}}]{pinto2012}
\bibinfo{author}{\bibfnamefont{P.~C.} \bibnamefont{Pinto}},
  \bibinfo{author}{\bibfnamefont{P.}~\bibnamefont{Thiran}}, \bibnamefont{and}
  \bibinfo{author}{\bibfnamefont{M.}~\bibnamefont{Vetterli}},
  \bibinfo{journal}{Physical Review Letters} \textbf{\bibinfo{volume}{109}},
  \bibinfo{pages}{068702+} (\bibinfo{year}{2012}),
  \urlprefix\url{http://dx.doi.org/10.1103/PhysRevLett.109.068702}.

\bibitem[{\citenamefont{Lokhov et~al.}(2014)\citenamefont{Lokhov, M\'ezard,
  Ohta, and Zdeborov\'a}}]{DMP_0}
\bibinfo{author}{\bibfnamefont{A.~Y.} \bibnamefont{Lokhov}},
  \bibinfo{author}{\bibfnamefont{M.}~\bibnamefont{M\'ezard}},
  \bibinfo{author}{\bibfnamefont{H.}~\bibnamefont{Ohta}}, \bibnamefont{and}
  \bibinfo{author}{\bibfnamefont{L.}~\bibnamefont{Zdeborov\'a}},
  \bibinfo{journal}{Phys. Rev. E} \textbf{\bibinfo{volume}{90}},
  \bibinfo{pages}{012801} (\bibinfo{year}{2014}),
  \urlprefix\url{http://link.aps.org/doi/10.1103/PhysRevE.90.012801}.

\bibitem[{\citenamefont{Altarelli et~al.}(2014)\citenamefont{Altarelli,
  Braunstein, {Dall{\textquoteright}Asta}, Lage-Castellanos, and
  Zecchina}}]{BBP}
\bibinfo{author}{\bibfnamefont{F.}~\bibnamefont{Altarelli}},
  \bibinfo{author}{\bibfnamefont{A.}~\bibnamefont{Braunstein}},
  \bibinfo{author}{\bibfnamefont{L.}~\bibnamefont{{Dall{\textquoteright}Asta}}},
  \bibinfo{author}{\bibfnamefont{A.}~\bibnamefont{Lage-Castellanos}},
  \bibnamefont{and} \bibinfo{author}{\bibfnamefont{R.}~\bibnamefont{Zecchina}},
  \bibinfo{journal}{Phys. Rev. Lett.} \textbf{\bibinfo{volume}{112}},
  \bibinfo{pages}{118701} (\bibinfo{year}{2014}),
  \urlprefix\url{http://link.aps.org/doi/10.1103/PhysRevLett.112.118701}.

\bibitem[{\citenamefont{Antulov-Fantulin
  et~al.}(2014{\natexlab{b}})\citenamefont{Antulov-Fantulin, Lancic, Stefancic,
  Sikic, and Smuc}}]{SourceDetectAvgTopK}
\bibinfo{author}{\bibfnamefont{N.}~\bibnamefont{Antulov-Fantulin}},
  \bibinfo{author}{\bibfnamefont{A.}~\bibnamefont{Lancic}},
  \bibinfo{author}{\bibfnamefont{H.}~\bibnamefont{Stefancic}},
  \bibinfo{author}{\bibfnamefont{M.}~\bibnamefont{Sikic}}, \bibnamefont{and}
  \bibinfo{author}{\bibfnamefont{T.}~\bibnamefont{Smuc}},
  \bibinfo{journal}{Proceedings of 2014 IEEE Eighth International Conference on
  Self-Adaptive and Self-Organizing Systems Workshops}  (\bibinfo{year}{2014}{\natexlab{b}}),
\urlprefix\url{http://dx.doi.org/10.1109/SASOW.2014.35}.

\bibitem[{\citenamefont{Comin and da~Fontoura~Costa}(2011)}]{SourceEigenCen_5}
\bibinfo{author}{\bibfnamefont{C.~H.} \bibnamefont{Comin}} \bibnamefont{and}
  \bibinfo{author}{\bibfnamefont{L.}~\bibnamefont{da~Fontoura~Costa}},
  \bibinfo{journal}{Phys. Rev. E} \textbf{\bibinfo{volume}{84}},
  \bibinfo{pages}{056105} (\bibinfo{year}{2011}),
  \urlprefix\url{http://link.aps.org/doi/10.1103/PhysRevE.84.056105}.

\bibitem[{\citenamefont{Brockmann and Helbing}(2013)}]{Brockmann}
\bibinfo{author}{\bibfnamefont{D.}~\bibnamefont{Brockmann}} \bibnamefont{and}
  \bibinfo{author}{\bibfnamefont{D.}~\bibnamefont{Helbing}},
  \bibinfo{journal}{Science} \textbf{\bibinfo{volume}{342}},
  \bibinfo{pages}{1337} (\bibinfo{year}{2013}), ISSN \bibinfo{issn}{1095-9203},
  \urlprefix\url{http://dx.doi.org/10.1126/science.1245200}.

\bibitem[{\citenamefont{Marron and Nolan}(1988)}]{KernelEstimation}
\bibinfo{author}{\bibfnamefont{J.~S.} \bibnamefont{Marron}} \bibnamefont{and}
  \bibinfo{author}{\bibfnamefont{D.}~\bibnamefont{Nolan}},
  \bibinfo{journal}{Statistics \& Probability Letters}
  \textbf{\bibinfo{volume}{7}}, \bibinfo{pages}{195} (\bibinfo{year}{1988}).

\bibitem[{\citenamefont{Rocha et~al.}(2011)\citenamefont{Rocha, Liljeros, and
  Holme}}]{RochaHolme}
\bibinfo{author}{\bibfnamefont{L.~E.~C.} \bibnamefont{Rocha}},
  \bibinfo{author}{\bibfnamefont{F.}~\bibnamefont{Liljeros}}, \bibnamefont{and}
  \bibinfo{author}{\bibfnamefont{P.}~\bibnamefont{Holme}},
  \bibinfo{journal}{PLoS Comput Biol} \textbf{\bibinfo{volume}{7}},
  \bibinfo{pages}{e1001109} (\bibinfo{year}{2011}),
  \urlprefix\url{http://dx.doi.org/10.1371%2Fjournal.pcbi.1001109}.

\bibitem[{\citenamefont{Korenromp et~al.}(2002)}]{GonnChly}
\bibinfo{author}{\bibfnamefont{E.~L.} \bibnamefont{Korenromp}}
  \bibnamefont{et~al.}, \bibinfo{journal}{Int J STD AIDS}
  \textbf{\bibinfo{volume}{13}}, \bibinfo{pages}{91} (\bibinfo{year}{2002}).



\end{thebibliography}

\end{document}